# Intelligent Decision Method for Main Control Parameters of Tunnel Boring Machine based on Multi-Objective Optimization of Excavation Efficiency and Cost

Bin Liu [a, b, c, *], Yaxu Wang [a, b], Guangzu Zhao [a, b], Bin Yang [d], Ruirui Wang [a], Dexiang Huang [d], Bin Xiang [d]

[a] *Geotechnical and Structural Engineering Research Center, Shandong University, Shandong, China*
[b] *School of Qilu Transportation, Shandong University, Shandong, China*
[c] *Data Science Institute, Shandong University, Shandong, China*
[d] *Zhejiang Tunnel Engineering Group Co., Ltd., Zhejiang, China*

**\* Corresponding author:** Bin Liu, corresponding at Geotechnical and Structural Engineering Research Center, Shandong University, No. 17923 Jingshi Road, Post 250061, Jinan, China.
**E-mail address:** liubin0635@163.com

**Abstract：**
　　Timely and reasonable matching of the control parameters and geological conditions of the rock mass in tunnel excavation is crucial for hard rock tunnel boring machines (TBMs). Therefore, this paper proposes an intelligent decision method for the main control parameters of the TBM based on the multi-objective optimization of excavation efficiency and cost. The main objectives of this method are to obtain the most important parameters of the rock mass and machine, determine the optimization objective, and establish the objective function. In this study, muck information was included as an important parameter in the traditional rock mass and machine parameter database. The rock–machine interaction model was established through an improved neural network algorithm. Using 250 sets of data collected in the field, the validity of the rock–machine interaction relationship model was verified. Then, taking the cost as the optimization objective, the cost calculation model related to tunneling and the cutter was obtained. Subsequently, combined with rock–machine interaction model, the objective function of control parameter optimization based on cost was established. Finally, a tunneling test was carried out at the engineering site, and the main TBM control parameters (thrust and torque) after the optimization decision were used to excavate the test section. Compared with the values in the section where the TBM operators relied on experience, the average penetration rate of the TBM increased by 11.10%, and the average cutter life increased by 15.62%. The results indicate that this method can play an effective role in TBM tunneling in the test section.
**Keywords:** tunnel boring machine tunneling; optimization and decision method of main control parameters; multi-objective optimization; field test

*List of symbols*

| | | | |
|---|---|---|---|
| C | Cost | PR | Penetration rate |
| $c_1$ | Cutter cost | Q | Quartz content |
| $c_2$ | Labor and material cost of the equipment | RQD | Rock quality designation |
| CAI | Cerchar abrasivity index | SA | Simulated annealing |
| CI | Coarseness index | SRC | Surrounding rock classification |
| $D_{TBM}$ | Cutterhead diameter | TBM | Tunnel boring machine |
| $E_f$ | Cutter life | Th | Thrust |
| L | Excavated tunnel length | Tor | Torque |
| M | Mean grain size | U | Utilization |
| MAE | Mean absolute error | UCS | Uniaxial compressive strength |
| MAPE | Mean absolute percentage error | W | Total wear of all cutters on the cutterhead |
| $M_{gt}$ | Muck geometry types | $W_{max}$ | Wear limit of cutter |
| CCR | Model relationship between cutter life and TBM control parameters and rock parameters | | |
| PRCR | Model relationship between penetration rate and TBM control parameters and rock parameters | | |

# 1. Introduction

For the past few years, tunnel boring machines (TBMs) have been widely used in tunnel excavations owing to their advantages of high tunneling efficiency, good safety, a high degree of mechanical automation, and informatization (Li et al., 2014, 2017; Liu et al., 2019a). Although substantial progress has been made in the field of tunnel construction, TBMs are restricted by many aspects in the process of TBM tunneling, including geological conditions of strata, machine design and operation parameters, and site construction management (Liu et al., 2017, 2018). Improper selection of control parameters or construction schemes often causes low tunneling efficiency, a sharp increase in the construction period and cost, and even machine blockage (Ma et al., 2009; Liu et al., 2016a). Several typical engineering cases also illustrate the challenges of TBM tunneling. For example, in the construction of the diversion tunnel of the Dul Hasti hydropower project in India, unsuitable TBM control parameters were selected for tunneling—due to an unclear understanding of the geological structure—which resulted in serious construction delays (Shang et al., 2007). During the excavation of the diversion tunnel of a hydropower station in Pakistan, the buried depth of the tunnel was 300–2000 m. Difficulty in adaptation of the control parameters of TBMs to high-strength surrounding rock may result in serious cutter wear and slow construction progress (Ma et al., 2018).

Currently, based on geological exploration data and excavation observations, operators adjust the control parameters according to their experience and operate manually within a rated value range. On the one hand, consideration of only the qualitative observation of geological changes results in a lack of diversified quantitative acquisition and limitation to analysis of key rock mass parameters. On the other hand, the selection of TBM control parameters depends mainly on experience, which is not based on different rock mass conditions, and there is no quantitative standard (Xue et al., 2018). If the geological conditions change suddenly or the operators are inexperienced, the control parameters and geological conditions of rock mass cannot be timeously and reasonably matched (Li et al., 2019). In this case, the selection of control parameters lacks a scientific and reasonable basis. This causes abnormal damage to the TBM, low rock-breaking efficiency, high cutting cutter consumption, and construction delays (Wang et al., 2018). Therefore, obtaining reasonable control parameters suitable for geological conditions has become a significant challenge for TBM tunneling.

To solve these problems, some meaningful work has been performed and reported in literature. Sun et al. (2016) developed a multidisciplinary design optimization (MDO) model for the design of TBMs. The model considers the construction period, energy consumption, and construction costs as the optimization objectives and uses the thrust, hydraulic power, and height–width ratio of the cutting ring as constraints to explore the optimal combination of the TBM structure and control parameters under different combination strategies. The results show that the proposed strategy with an adaptive structure and control parameters can significantly reduce construction period and energy consumption. Based on the MDO model established by Sun et al. (2016), Wang et al. (2018) proposed a reliability-based performance optimization strategy by considering the stochastic properties of rocks—the performance of the TBM could be optimized according to the changing geological conditions with uncertainties. Xue et al. (2018) analyzed the energy relationship in the process of tunneling. According to the law, the best tunneling specific energy and corresponding optimum operating parameters (thrust and torque) could be obtained, providing an evaluation standard for the selection of TBM operation parameters. Li et al. (2019) established an information and intelligence technology application system for TBMs. Based on the learning of historical tunneling data and operator behavior, the idea of intelligent control of TBMs in general and adverse formations was proposed. Through real-time monitoring of rock mass and machine parameters, combined with operating behavior, the control and optimization of TBM tunneling was realized. Tan (2020) acquired cutterhead revolutions per minutes, thrust, and other parameters in real time during TBM tunneling and filtered the invalid tunneling information through a set boundary range. The effective information could then be used to predict the control parameters in real time and provide parameter suggestions for operators. Current research has made an effective progress and contributed to intelligent TBM tunneling. At present, the optimization of TBM control parameters is performed in two ways: historical data mining and objective optimization. The former trains an effective model through learning a large amount of historical data and then realizes the decision of control parameters based on the similarity of geological conditions. This method depends on the quality of historical data and the proportion of excellent schemes. Conversely, the principle of the multi-objective optimization method is to find the relationship among the main control parameters, tunneling performance, and rock mass parameters. On this basis, the control parameters corresponding to the optimal tunneling performance can be obtained. This method has less dependence on historical data and uses the current optimal state as the decision condition. Consequently, it has good research prospects and application potential.

The key to multi-objective optimization lies in three aspects. Firstly, what are the relevant rock parameters and machine parameters to be used. Secondly, which optimization objectives should be considered as the standard. Thirdly, how does one establish the objective function of the optimization decision. Regarding the first aspect, for the selection of the TBM machine parameters and rock parameters, the rock mass parameters commonly used include rock mass strength and integrity. Machine parameters commonly used include thrust, torque, and penetration. Meanwhile, some researchers have explored the impact of muck on TBM tunneling. Heydari et al. (2019) studied the relationship between various TBM operational factors, performance, and muck geometry. The analysis results show that there is a strong inverse correlation between specific energy and muck size indicators. Through the simulation of the cutterhead rock-breaking process, Geng et al. (2019) found that the shape of the muck had a great influence on TBM tunneling. The muck geometry played an important role in assisting operators in judging the integrity of the surrounding rock and the wear condition of the cutter. Therefore, muck information should be included in the multi-objective optimization as a type of information representing the rock–machine interaction state. For the second aspect, most previous studies used construction period, energy, and construction costs as objectives, which could be used to make optimal decisions (Sun et al., 2016; Xue et al., 2018). For the third aspect, firstly, in the establishment of rock–machine mapping, some researchers have established rock–machine interaction relationship from the physical rule of rock breaking using a cutter through linear cutting test (Cho et al., 2013; Gong et al., 2007; Rostami et al., 1996). Others have used artificial intelligence algorithms to establish rock–machine interaction relationships by collecting a large amount of data in engineering (Liu et al., 2019b, 2020c; Salimi et al., 2018). Secondly, by using rock–machine mapping and analyzing the relationship between optimization objectives and decision parameters, the objective function could be established.

The tunneling efficiency and cutter wear of a TBM in the construction process are the main factors affecting the construction period and cost. Evidently, the speed of the TBM excavation directly affects the construction period of the project. In some construction cases of TBM tunneling, the time of inspection, replacement, and maintenance of disc cutters can take up approximately 30% of the construction period (Yang et al., 2019). The longer the construction period, the higher the input of various costs. Meanwhile, the cost of disc cutter replacement contributes approximately one-third of the total construction cost (Liu et al., 2017c; Sun et al., 2019; Tian et al., 2020). These are important factors that need to be considered by TBM operators to adjust control parameters (Chung et al. 2019; Faramarzi et al., 2020; Noori et al., 2020). Therefore, the penetration rate (PR) and cutter life ($E_f$) are two key parameters in the cost analysis used in this study.

The primary idea underpinning this study can be expressed as follows: muck information is included as an important parameter in the traditional rock mass and machine parameter database. The costs related to the PR and $E_f$ are considered, and the cost is considered as the optimization goal. The rock–machine interaction model is established through an improved neural network algorithm. On this basis, the objective function of optimal decision making can be established. The research contents of this paper can be summarized as follows: the optimized back propagation (BP) neural network algorithm is used to mine and analyze a large number of rock masses and the TBM machine data. The rock–machine interaction relationship models of the TBM PR and $E_f$ are established, and the objective function is instituted accordingly. Based on the optimal cost, a control parameter decision method for a TBM is proposed to determine the control parameters that match the geological conditions, so as to improve the tunneling efficiency of the TBM and reduce the costs. The rest of this paper is organized as follows. Section 2 introduces the main ideas of the TBM control parameter decision-making method.

The engineering background, basic TBM information, and data sources of rock mass and TBM control parameters are provided in Section 3. The establishment of related models in the decision-making method and the specific expression of the method are introduced in Sections 4 and 5. Section 6 describes an excavation experiment and presents a discussion using decision methods. Finally, the conclusions and research perspectives are provided in Section 7.

## 2. Decision method of main control parameters

In normal formations without significant adverse geology, the goal of TBM tunneling is high efficiency and low cost. In principle, there are two key points to realize using this multi-objective optimization method. The first is how to establish the objective function. The second is how to find the global optimal solution. The premise of establishing the objective function is to obtain the corresponding rock–machine interaction relationship. Specifically, that is to establish the relationship model between the PR and TBM control parameters and rock parameters (PRCR model), and the $E_f$ between the TBM control parameters and rock parameters (CCR model), respectively. Then, the objective function of cost, including the PR and $E_f$, is constructed. The corresponding optimal TBM control parameters can be obtained through global optimization. In this study, according to the historical data of the TBM control parameters in practical projects, their variation ranges were determined. Set a certain step size to traverse all the data in this range to obtain the optimal solution. These parameters are reasonable and scientific control parameters acquired through the decision process. The implementation flowchart is shown in Fig. 1.

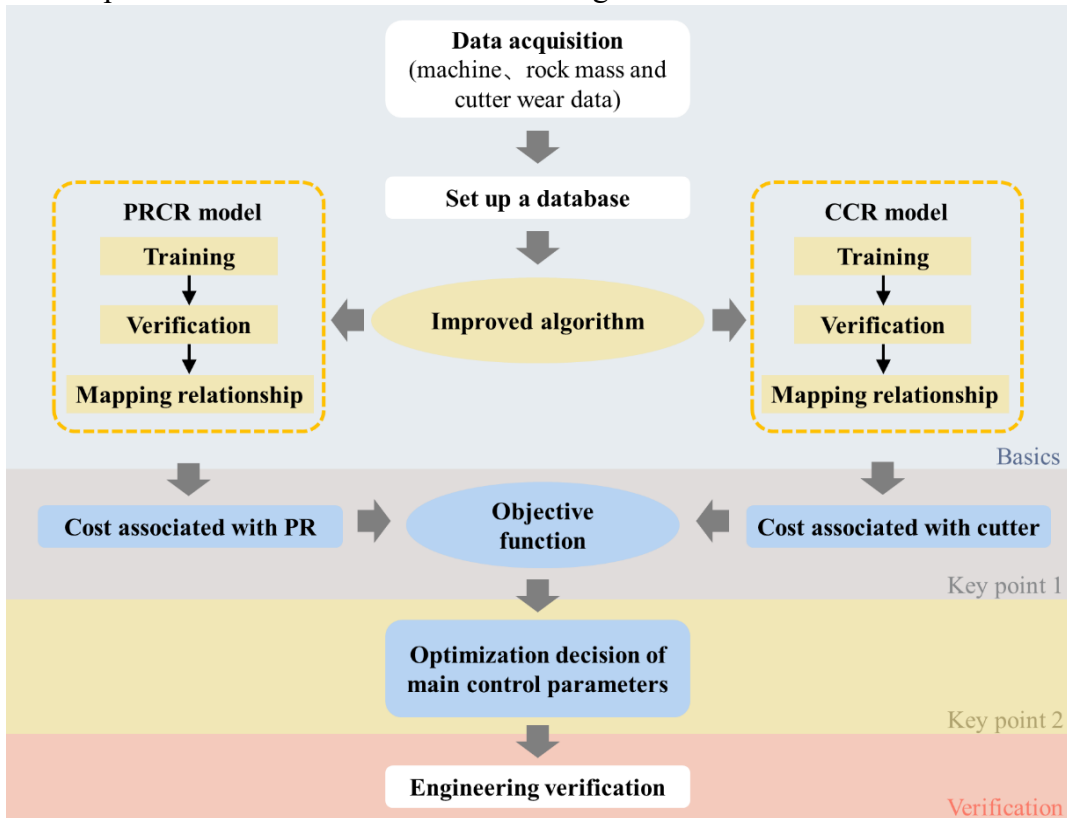

**Fig. 1** Flow chart of decision making for optimization of main control parameters.

First, the objective function was established. In this study, the objective function of the optimization of the TBM main control parameters optimization is to minimize the cost. As in the previous analysis, the maintenance and replacement of worn cutters costs time and increases the cost of cutters. At the same time, the PR is closely related to the construction period, and any increase in construction period will increase many construction costs. In other words, typically, cost is positively related to the construction period and negatively related to the tunneling speed. Therefore, in the process of establishing the objective function, the related cost of the PR and cutters are primarily considered. Furthermore, a few aspects of the TBM construction cost is unrelated to the construction period, such as lining support, assembling segments, and other engineering materials costs, which are related to the tunnel quantity and tunnel length. These costs are not included in the cost model established in this study. Using the above analysis, the cost ($C$) calculation model including disc cutters cost ($C_1$) and construction period-related cost ($C_2$) is established, as shown in Eq. (1).

$$C=C_1+C_2 \tag{1}$$

Second, the global optimal solution was determined. Similarly, the premise of this step was to use the previously proposed PRCR and CCR models. When the rock mass parameters are known and the control parameters change in a given range, the PR and $E_f$ corresponding to the control parameters can be obtained through the aforementioned two models. Subsequently, all the costs within the range of control parameters were calculated by the cost model including PR and $E_f$; an exhaustive search was conducted to find the optimal cost. Thus, the optimal decision value was obtained with the control parameters under the optimal cost.

Because the rock–machine interaction relationship model (PRCR and CCR model) was a basis for the use of decision methods, to explain the establishment and use of the model in more detail, a specific project and data were employed for illustration.

## 3. Project profile

### 3.1 Project description

The project supporting this paper is the Shanling Section of the Hangzhou Second Water Source Water Transfer Channel Project (Jiangnan Line). The route diagram of the Shanling Section is shown in Fig. 2. The tunnel length of this section was 13.026 km. The excavation scheme of combining double-shield TBM with drilling and blasting method was adopted. In Fig. 2, the TBM tunneling section is indicated by the red line, representing a distance of approximately 8.036 km. The Shanling Section of the tunnel mainly comprised low and medium mountains and hills, with large surface relief. The elevation of the highest point along the tunnel was approximately 390 m. According to the geotechnical engineering investigation report, the project section was distributed in an east-west direction, approximately perpendicular to the tunnel. The formation lithology was mainly composed of medium and fine-grained sandstone intercalated with silty mudstone, argillaceous siltstone, and siltstone. Joint fissures were developed with a steep dip angle. The joint surface was mainly filled with thin calcite, and the overall integrity of the rock mass was poor. The strength of the surrounding rock should have been distributed between 30 MPa and 150 MPa. The surrounding rock was classified based on the Hydropower Classification (HC) method of national standards—most of them being III or IV of the surrounding rock classification, accounting for 38% and 52% of the tunnel length, respectively—and there were several fault fracture zones. The CREC696 double-shield TBM produced by the China Railway Engineering Equipment Group was selected for the construction. The cutterhead was 6 m in diameter and equipped with 34 cutters, including sizes of 17 in and 19 in. The cutterhead diameter is an important parameter used to establish the optimization and decision model of the TBM control parameters. Tunnel excavation, segment installation, pea gravel and mortar backfill, slurry mixing, and grouting were all completed using the double-shield TBM equipment. Descriptions of the tunnel and TBM's main technical parameters are shown in Table 1.

**Table 1**
Main technical parameters of CREC696 TBM.

| Parameters name | Values | Unit |
| --- | --- | --- |
| TBM type | Double shield | - |
| Cutterhead diameter | 6000 | mm |
| Cutterhead rotation speed | 0~5.3~10.19 | rpm |
| Maximum PR | 120 | mm/min |
| Cutter size | 17, 19 | inch |
| Number of cutters | 38 | piece |
| Rated torque | 3780 | KN·m |
| Rated thrust | 3000 | T |
| Power | 3780 | KW |

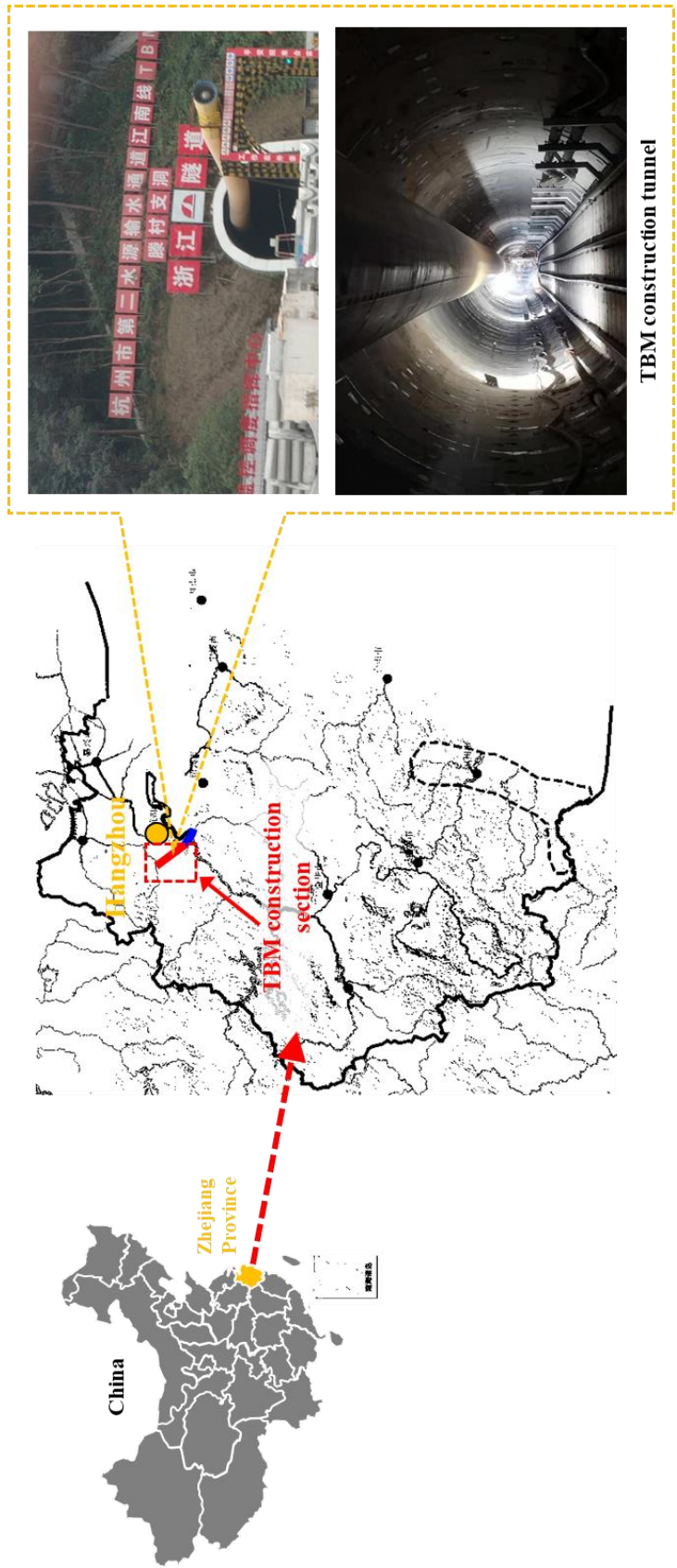

**Fig. 2** The route diagram of the Shanling Section of the Hangzhou Second Water Source Water Transfer Channel Project (Jiangnan Line).

## 3.2 Dataset

The acquisition of relevant machine and rock parameters is particularly important for optimizing the TBM control parameter. Consequently, some main rock and machine parameters were selected to explore the rock–machine interaction relationship of the TBM and provide data support for the optimization decision of the TBM control parameters.

### 3.2.1 TBM machine and tunneling parameters

The TBM machine and tunneling parameters used in this study include thrust (Th), torque (Tor), PR, and $E_f$. Among them, PR is directly related to the tunneling time of the TBM, which is a crucial performance parameter for TBM operators. In the process of TBM tunneling, the thrust and torque of the cutterhead directly affect the rock-breaking effect of the TBM cutters and are also the key factors affecting disc cutter wear (Su et al., 2020; Wang et al., 2020b). The thrust of the TBM is directly related to the normal force in the process of rock breaking using the disc cutter. In addition, the normal force determines whether the disc cutter can efficiently penetrate the rock surface. The torque of the TBM is directly related to the rolling force of the disc cutter. The two main forces are the main driving factors for crack initiation and rock spalling (Pan et al., 2019a, 2019b). Therefore, thrust and torque are very significant control parameters in TBM tunneling; consequently, the aforementioned target parameters need to be optimized.

As one of the important working parts of a TBM, disc cutters directly contact the surrounding rock. The factors affecting the degree of disc cutter wear mainly include the machine parameters and rock mass conditions of the TBM. In general, the higher the integrity and compressive strength of the rock mass, the higher the wear of the TBM disc cutters (Jain and Lad, 2015; Wang, et al. 2020a). Moreover, the life of the cutter is also affected by control parameters such as thrust and torque. In particular, under the action of high torque and thrust, abnormal wear occurs (Wang et al., 2020b). The cutter life can be characterized by the tunneling distance of each cutter and volume of the surrounding rock excavated. It can also be the volume of rock excavated per 1 mm of cutter ring wear. In this study, the data of the daily excavation length and amount of wear were obtained through the cutter wear records of the TBM engineering site. Subsequently, the TBM cutter life was calculated using Eq. (2) (Liu et al., 2017c).

$$E_f(\text{m}^3/\text{mm}) = \frac{\pi \cdot D^2_{TBM} \cdot l}{4 \cdot W} \tag{2}$$

where $W$ is the total wear extent of all disc cutters used on the cutterhead (mm); $D_{TBM}$ is the cutterhead diameter (m); and $l$ is the excavation distance corresponding to total tool wear (m). $E_f$ is a general index used to evaluate the cutter life.

Therefore, the following four TBM machine parameters were selected: PR, Th, Tor, and $E_f$. The CREC696 TBM has a tunneling data-recording platform. The control, performance, hydraulic parameters, and other parameters of the TBM are recorded on the platform at a frequency of one per second. Through the data platform, the relevant parameters and mileage data of the TBM excavation could be obtained.

### 3.2.2 Rock mass parameters

As an important part of construction, geological work needs to be carried out at the same time in the TBM tunneling process. The data of the rock mass parameters used in this study were from the 7+650 to 5+550 mileage section of the project. The rock mass parameters included SRC, UCS, RQD, Cerchar abrasivity index (CAI), quartz content (q), and muck information. Among them, SCR, UCS, RQD, CAI, and q are the representative and important rock mass parameters that researchers have focused on in the previous studies (Armaghani et al., 2019; Armetti et al., 2018; Mohammadi et al., 2015; Macias et al., 2016; Liu et al., 2016b; Ribacchi and Fazio, 2005; Su et al., 2020; Suana and Peters, 1982). Through drilling cores in tunnels and machining them into standard specimens. The aforementioned parameters are obtained by standard mechanical tests. Part of the UCS is also obtained by point load test. The muck information is also an important parameter for obtaining the TBM tunneling state. For example, Cho et al. (2013), after performing the linear cutting test, found that under specific surrounding rock conditions, the change in control parameters led to a change in muck morphology. Mohammadi et al. (2019) also found muck information to be helpful in studying the rock-breaking efficiency, mechanical properties, and rock-breaking mechanism of the TBM. Therefore, in this study, muck size, coarseness index (CI), and mean grain size (M) were selected as three representative parameters of muck information.

Based on existing research on muck geometry types ($M_{gt}$) (Heydari et al., 2019), and combined with the actual construction site situation, the muck type was divided into three types: rock debris, rock slices, and rock blocks (Zhao, 2020). Using a monitoring video installed on the TBM conveyor belt and combined with the construction log (Fig. 3), rock dregs of different forms were combined and classified into four categories for calibration, as shown in Table 2. Consequently, there were four types of muck combinations. CI is defined as a comparative size distribution of the cut rocks and has been used in many studies. The value of CI is the sum of the accumulated weight percentage retained in each layer of screen during the screening test of the muck. More specifically, the proportion of the muck weight of each layer of sieve in the total weight can be calculated after sieving. Subsequently, the cumulative percentage of each layer of sieve can be obtained by successively accumulating. Further, the sum of the cumulative percentages of each layer of sieve is equal to CI. As shown in Eq. (3):

$$\text{CI} = \sum_{i}^{n} \sum_{i=1}^{n} w_i \tag{3}$$

where $n$ is the number of layers of sieves used, and $w_i$ is the residue mass of the $i$-th sieve. CI is related to the tools used in screening experiments; the results obtained by selecting different levels of sieves were not the same. In this study, six grades of sieves were selected—2.36, 4.75, 9.5, 19, 37.5 and 63 mm. The mean grain size (M) is a widely used classification method for describing rock particle size. There are several formulas for calculating the average particle size; however, the most commonly used method is the one as shown in Eq. (4) (Heydari et al., 2019). After sieving, the cumulative mass proportion of the muck in each particle size grade was measured. The average long-axis particle size of muck with mass ratios of 16%, 50% and 84% were calculated and this value is equal to M.

$$M = \frac{\phi 16 + \phi 50 + \phi 84}{3} \tag{4}$$

**Table 2**
Different types of muck geometry.

| $M_{gt}$ | Types | | |
| --- | --- | --- | --- |
| | Rock debris | Rock slices | Rock block |
| 1 | √ | | |
| 2 | √ | √ | |
| 3 | √ | | √ |

|   |   |   |   |   |
|---|---|---|---|---|
| 4 | √ |   | √ |   | √ |

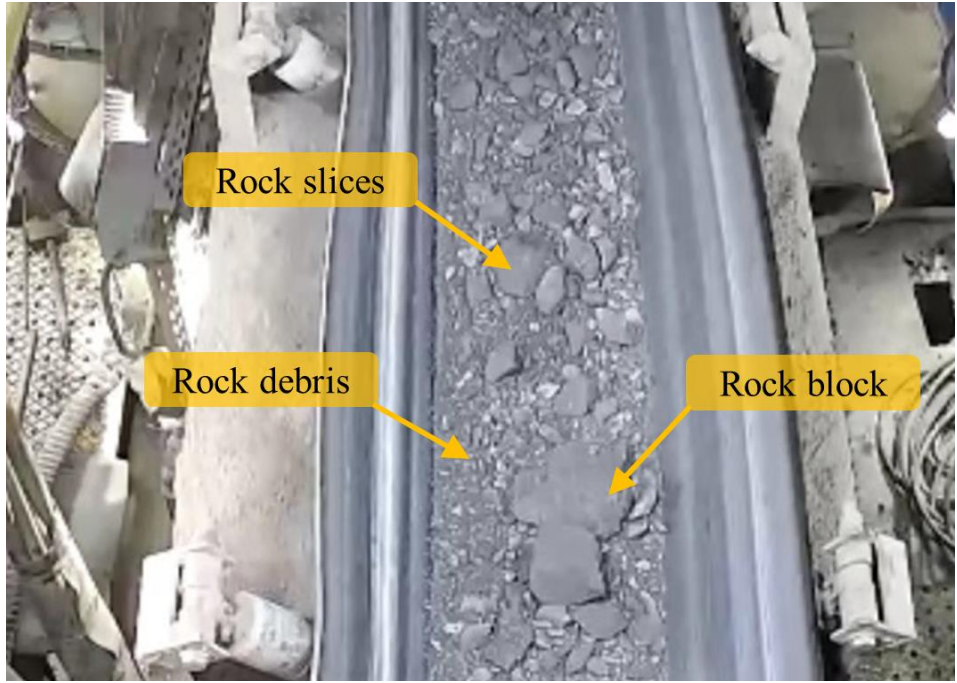

**Fig. 3** Conveyor belt monitoring video and muck shape calibration.

## 4. Rock–machine interaction relationship model

In this study, a simulated annealing (SA) improved BP neural network (BPNN) algorithm was used to establish the rock–machine interaction relationship model. Neural networks have been widely used in several fields owing to their excellent self-learning and generalization abilities (Li et al., 2020; Liu et al., 2020a; Liu et al., 2020b). It is also popular in mining the TBM rock–machine interaction relationship (Armaghani et al., 2017; Fattahi and Bazdar, 2017; Koopialipoor et al., 2020; Wei et al., 2020). The BPNN is a widely used neural network, the back error propagation is used to update the current network weight and threshold using the gradient descent method. Subsequently, the optimal value is output through the network. However, the gradient descent method often falls into a local extreme value and cannot achieve a good convergence effect. The SA algorithm is a heuristic global search method that is very effective for solving nonlinear problems with multiple local optimizations. Therefore, in this study, the SA-BPNN algorithm proposed by Liu et al. (2020c) was used to establish the rock–machine interaction relationship model of PR and $E_f$. The calculation process of this method is as follows. First, a group of weights and biases of the BPNN is generated randomly. The weights and biases are updated using the SA algorithm. The values obtained are the initial weights and biases of the BPNN. Finally, the ultimate model is obtained by training the BPNN.

While establishing the neural network model, first, the input and output parameters should be determined. The input parameters used to build the model include eight rock mass parameters—SRC, UCS, RQD, CI, M, $M_{gt}$, CAI, and q—and two machine parameters—Th and Tor. The output parameters are PR and $E_f$. The input and output relationships corresponding to the rock–machine interaction are shown in Fig. 4. The two models select the same rock mass and machine parameters as the input, in order to more effectively use the existing data to fully mine and learn the relationship between these parameters and establish a better model mapping relationship. Because there are category data in the input data, it is necessary to deal with discontinuous numerical features using one-hot encoding. For example, if the $M_{gt}$ is 1 and 3, after code conversion, it can be expressed as [1,0,0,0] and [0,0,1,0].

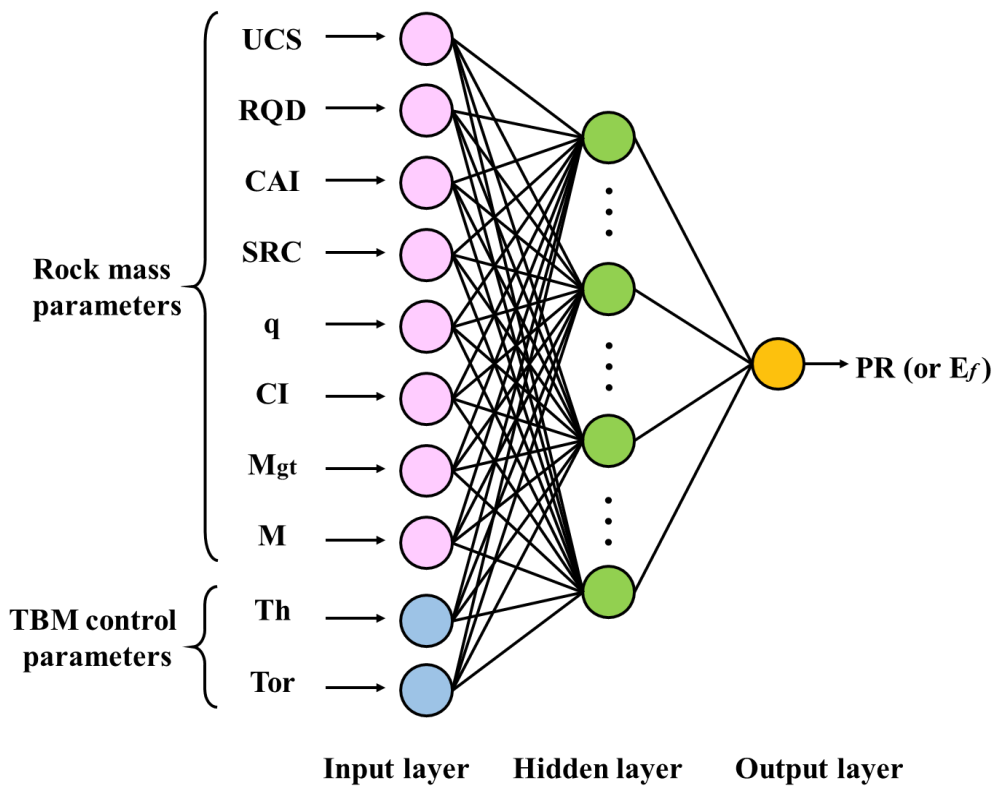

**Fig. 4** The input and output relations corresponding to rock-machine interaction.

In order to fully train the model and improve the generalization ability of the model. In this study, using 160 rock parameters and machine parameter samples to mine the rock–machine interaction relationship of the PRCR model. In addition, another 90 samples were used to mine the rock-mechanical interaction relationship of the CCR model. The statistical information of parameters related to the two models are shown in Table 3 and Table 4. The proportion of different lithology, surrounding rock classification, and muck geometry types in the data were shown in Fig. 5 and Fig. 6. For the input parameters of the neural network, the correlation between the other parameters is analyzed except for the category data. The correlation coefficients ($r$) between them are shown in Table 5. It can be seen that the correlation coefficient between UCS and Th is 0.729, and that between CI and M is 0.835. There is a correlation between the two groups of coefficients. The correlation between other parameters is not significant. To overcome the autocorrelation of the two groups of parameters, principal component analysis (PCA) is used to extract the common features of the two related feature parameters. The extracted data is taken as the input feature. In addition, the z-score normalization method was used to process all the data to eliminate the influence of dimension and order of magnitude.

**Table 3**

Statistical information of PRCR model parameters.

| Parameters | Mean | Maximum | Minimum | Std. deviation |
| --- | --- | --- | --- | --- |
| **UCS (MPa)** | 70.97 | 149.03 | 30.35 | 32.57 |
| **RQD (%)** | 45.84 | 93.01 | 5.00 | 22.91 |
| **CAI** | 4.28 | 5.32 | 2.13 | 0.53 |
| **q (%)** | 75.09 | 95.21 | 50.38 | 9.89 |
| **CI** | 374.39 | 590.30 | 257.09 | 60.95 |
| **M (mm)** | 14.38 | 30.47 | 1.35 | 6.64 |
| **Th (KN)** | 4952.66 | 9127.08 | 2105.16 | 1947.83 |
| **Tor (KN·m)** | 782.96 | 1327.25 | 222.49 | 348.87 |
| **PR (mm/min)** | 56.82 | 95.55 | 25.89 | 16.87 |

**Table 4**
Statistical information of CCR model parameters.

| Parameters | Mean | Maximum | Minimum | Std. deviation |
| --- | --- | --- | --- | --- |
| **UCS (MPa)** | 70.80 | 149.03 | 36.81 | 28.01 |
| **RQD (%)** | 42.02 | 90.35 | 6.52 | 20.43 |
| **CAI** | 3.16 | 4.52 | 2.12 | 0.47 |
| **q (%)** | 71.10 | 93.40 | 43.82 | 11.35 |
| **CI** | 281.89 | 507.77 | 229.78 | 56.91 |
| **M (mm)** | 12.24 | 36.24 | 1.78 | 6.74 |
| **Th (KN)** | 5495.85 | 9127.08 | 2543.61 | 1818.15 |
| **Tor (KN·m)** | 805.29 | 1281.82 | 245.97 | 332.69 |
| **E$_f$ (mm/m$^3$)** | 33.67 | 56.00 | 15.00 | 8.62 |

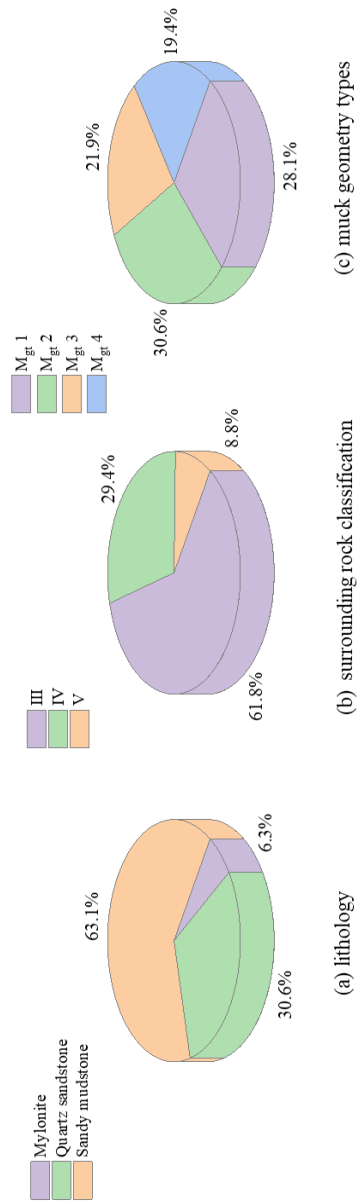

**Fig.5** The proportion of lithology, surrounding rock classification and muck geometry types in PRCR model data.

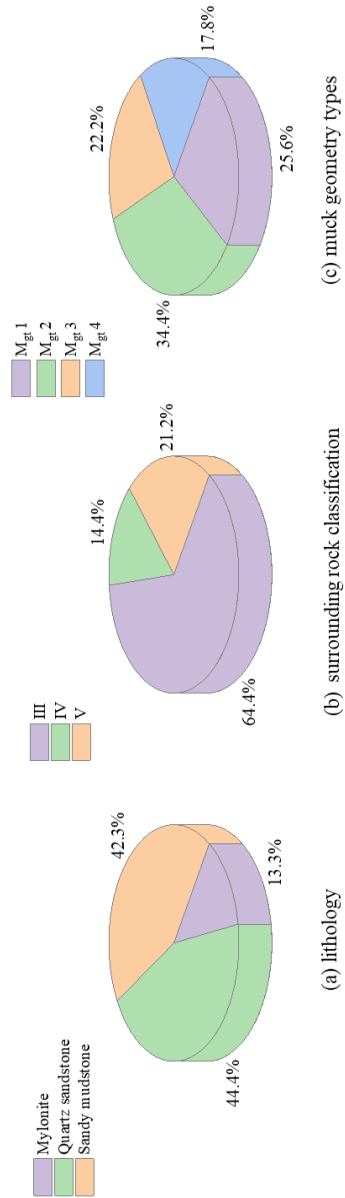

**Fig.6** The proportion of lithology, surrounding rock classification and muck geometry types in CCR model data.

**Table 5**
Correlation coefficient between the input parameters of the model.

| r | UCS | RQD | CAI | q | CI | M | Th | Tor |
|---|---|---|---|---|---|---|---|---|
| **UCS** | 1 | 0.187 | 0.321 | 0.201 | 0.073 | 0.064 | 0.729 | 0.444 |
| **RQD** | 0.187 | 1 | -0.031 | 0.199 | -0.136 | -0.105 | -0.008 | 0.056 |
| **CAI** | 0.321 | -0.031 | 1 | 0.551 | -0.003 | -0.145 | 0.450 | 0.339 |
| **q** | 0.201 | 0.199 | 0.551 | 1 | -0.051 | -0.123 | 0.344 | 0.200 |
| **CI** | 0.073 | -0.136 | -0.003 | -0.051 | 1 | 0.835 | 0.080 | 0.045 |
| M | 0.064 | -0.105 | -0.145 | -0.123 | 0.835 | 1 | 0.056 | -0.011 |
| Th | 0.729 | -0.008 | 0.450 | 0.344 | 0.080 | 0.056 | 1 | 0.593 |
| Tor | 0.444 | 0.056 | 0.339 | 0.200 | 0.045 | -0.011 | 0.593 | 1 |

Furthermore, it is necessary to set the network parameters, which directly affect the accuracy of the model. Network parameters can be divided into normal parameters and hyperparameters. The normal parameters can be updated using the gradient descent method (weight and bias). The hyperparameters include the number of network layers, number of network nodes, number of iterations, and learning rate. The parameters that need to be set in the SA algorithm are initial temperature, temperature drop ratio, and internal cycle times. The final parameter values determined after several tests are shown in Table 6 and Table 7. Simultaneously, for the PR and $E_f$ relationship models, 3-fold cross and 4-fold cross were used to verify the model, respectively. Finally, additional 40 groups of data and 18 groups of data were used to test the optimal PRCR model and CCR model, respectively.

**Table 6**
Setting and values of hyperparameters of PRCR model.

| Algorithms | Name | Set results |
|---|---|---|
| BP neural network | Algorithm | Gradient descent |
| | Learn rate | 0.1 |
| | Iterations | 2000 |
| | Network nodes | 11 |
| | Activation function | tanh |
| Simulated annealing (SA) | Initial temperature | 100 |
| | Temperature drop ratio | 0.99 |
| | Internal cycle times | 50 |
| | Final temperature | 0 |
| | Iterations | 1000 |

**Table 7**
Setting and values of hyperparameters of the CCR model.

| Algorithms | Name | Set results |
|---|---|---|
| BP neural network | Algorithm | Gradient descent |
| | Learn rate | 0.15 |
| | Iterations | 1000 |
|  | Network nodes | 12 |
|  | Activation function | tanh |
| **Simulated annealing (SA)** | Initial temperature | 80 |
|  | Temperature drop ratio | 0.99 |
|  | Internal cycle times | 30 |
|  | Final temperature | 0 |
|  | Iterations | 1000 |

    While establishing the PRCR model, first, the accuracy of the model was compared by calculating the three validation sets. By comparing the calculated value with the measured value, the calculation accuracy of the PR model can be directly expressed. To better evaluate the rock–machine interaction relationship model, the mean absolute error (MAE) and mean absolute percentage error (MAPE) of the calculated values were calculated. The absolute error reflects the accuracy of the model to a certain extent and can evaluate the overall effect of the model more objectively. The MAE and MAPE of the three validation sets were 6.56 mm/min, 6.52 mm/min, and 7.99 mm/min, and 12.98%, 12.60%, and 14.56%, respectively. Through calculation and comparison, it can be seen that the fitting effect with real value of the verification set 2 was the best among the three subsets. The results of using the remaining 40 groups of data as a test set to test the rock–machine interaction relationship model are shown in Fig. 7. The trend accuracy of the SA-BPNN model is more than 80% (the ratio of samples with the same change trend as the measured value to the total number of samples). The calculation effect was also achieved in the test set. The MAE and MAPE values were 8.89 mm/min and 13.14%, respectively, as shown in Table 8. Liu et al. (2020c) confirmed that the calculation accuracy of the SA-BPNN algorithm was better than that of the traditional BP neural network. The values calculated using the BPNN algorithm are also shown in the figure. The rock–machine interaction model based on the SA-BPNN algorithm exhibited higher accuracy.

    While establishing the CCR model, through the verification of the four verification sets, it could be concluded that the calculated and measured values were well fitted. Similarly, the MAE and MAPE values were calculated to measure the prediction effect of the model. As shown in Table 9, the MAE values were 3.86 m$^3$/mm, 3.89 m$^3$/mm, 4.52 m$^3$/mm, and 4.53 m$^3$/mm, and the MAPE values were 12.57%, 13.16%, 13.27%, and 12.95%. The model of verification set 1 was better than the other three models; thus, it was selected to calculate the test set data. The results using the remaining 18 groups of data as a test set to test the relationship model, are shown in Fig. 8. In the test set, the changing trend was essentially the same as the measured value, and it could also achieve high accuracy. Among them, 88.89% of the calculated values had the same trends as the measured values. The MAE and MAPE were 4.74 m$^3$/mm and 14.06%, respectively. In addition, the calculation results of the BPNN model are compared in Fig. 8; the SA-BPNN algorithm exhibits higher accuracy. The two rock–machine interaction relationship models established in this study could sufficiently represent the relationship between the input and output parameters, which could meet the practical engineering application requirements and exhibit a good generalization ability. It provides a relationship between the PR, $E_f$,

and rock mass parameters and control parameters of the TBM. simultaneously, it also lays the foundation for further research on the main control parameter decision method.

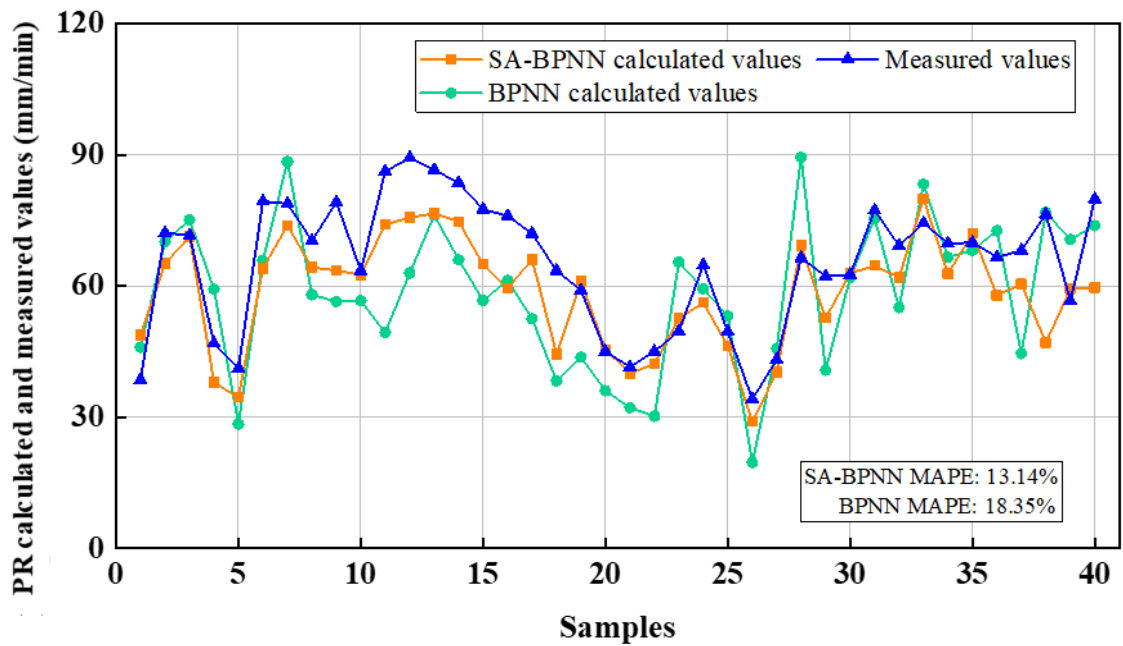

**Fig. 7** Comparison of calculated values and measured values of PR rock-machine interaction relationship model on test set.

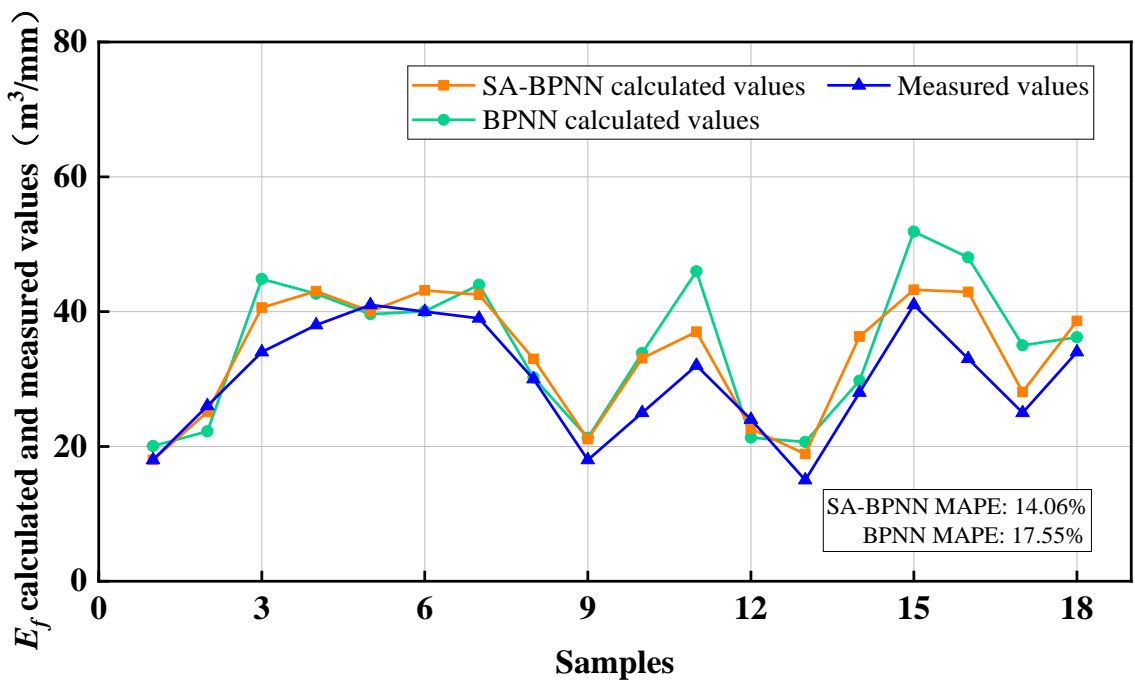

**Fig. 8** Comparison of calculated values and measured values of $E_f$ prediction model on test set.

**Table 8**
Evaluation index of PR rock–machine interaction relationship model with SA-BPNN.

| Data set | Evaluation index | |
|---|---|---|
| | MAE (mm/min) | MAPE (%) |
| Verification set 1 | 6.56 | 12.98 |
| Verification set 2 | 6.52 | 12.60 |
| Verification set 3 | 7.99 | 14.56 |
| Test set | 8.89 | 13.14 |

**Table 9**

Evaluation index of $E_f$ rock–machine interaction relationship model with SA-BPNN.

| Data set | Evaluation index | |
|---|---|---|
| | MAE (m³/mm) | MAPE (%) |
| Verification set 1 | 3.86 | 12.57 |
| Verification set 2 | 3.89 | 13.16 |
| Verification set 3 | 4.52 | 13.27 |
| Verification set 4 | 4.53 | 12.95 |
| Test set | 4.74 | 14.06 |

## 5. Specific expression of decision methods

The optimization and decision ideas similar to those in this study have been preliminarily explored by our team in previous study (Zhao, 2020). However, in this study, we have considered the influence of more types of rock mass parameters (including CI and M). Therefore, a new rock–machine interaction relationship model was established, and the objective function was improved, as shown in Eq. (5), which was a complete optimization decision process.

$$\begin{cases} C = min[c_1(\frac{\pi \cdot L \cdot D_{TBM}^2}{4 \cdot E_f \cdot W_{max}}) + c_2(\frac{L}{PR \cdot T})] \\ PR = f_1(x_1, x_2, x_3, x_4, x_5, x_6, x_7, x_8, Th, Tor) \\ E_f = f_2(x_1, x_2, x_3, x_4, x_5, x_6, x_7, x_8, Th, Tor) \\ Th_{min} < Th < Th_{max} \\ Tor_{min} < Tor < Tor_{max} \end{cases} \quad (5)$$

where $C$ is the cost, comprising cutter consumption-related cost and tunneling-related cost. $C_1$ is the cutter cost coefficient (the unit being RMB per cutter), which mainly includes the cost of purchasing cutters and maintenance. $C_2$ is the labor and material cost coefficient of the equipment (the unit being RMB per day), which mainly includes the labor cost, board and lodging cost paid to the workers in TBM tunneling, as well as the use, transportation, and installation cost of on-site construction equipment. According to the real cost recorded in the construction management of the project, the average daily cost is calculated as the value of the two parameters. While calculating the cost related to cutter consumption, it is assumed that the cutters wear in normal form during the process of tunneling. The ratio of the total amount of cutter wear corresponding to the excavation volume of the whole tunnel to the limit value of cutter wear is the total consumption of cutters, where $W_{max}$ is the wear limit of the cutter, which is 25 mm. In this study, the construction period is expressed as the ratio of the tunnel length to the actual tunneling distance per day, where $T$ is the normal effective tunneling time of the TBM per day and $L$ is the tunnel length of the TBM. To make the formula not only applicable to a certain project, $L$ is set to a unit length of 1 m. While calculating the cost related to tunneling, to simplify the calculation, the tunneling time was set as a fixed value. In this project, the daily tunneling time of TBM was relatively fixed. There was a forced downtime for maintenance of 4 h. According to the data recorded on site, the tunneling time varies from 8.7 to 10.8 h per day. Therefore, the daily tunneling time was set to 10 h. In this way, the cost calculation model was obtained, as shown in the first equation of Eq. (5).

The remaining equations in Eq. (5) can be explained as follows: $f_1$ is the PRCR model; $f_2$ is the CCR model; $x_1$ is the UCS; $x_2$ is the CI; $x_3$ is the muck geometry; $x_4$ is the M, $x_5$ is the SRC; $x_6$ is the RQD; $x_7$ is the CAI; and $x_8$ is the q. $Th_{min}$ and $Th_{max}$ are the thrust preset interval limit values. $Tor_{min}$ and $Tor_{max}$ are the torque preset interval limit values. When the rock mass condition is known, the thrust and torque changes in the preset interval corresponding to different PR and $E_f$. By changing the thrust and torque, we can determine the increase and decrease in cost under each working condition. Specifically, the definition domain is determined according to the historical data of the control parameters. Multiple sets of thrust and torque data were obtained by setting the step size. The thrust and torque values corresponding to the minimum cost were then obtained by global optimization, and these were the optimal main control parameters under the cost optimization objective.

# 6. Results and discussion

To verify the effectiveness of the optimization decision method, a section was selected for the tunneling test in the Shanling Section of the Hangzhou Second Water Source Water Transfer Channel Project (Jiangnan Line). The selected verification section had to avoid the existence of faults and other unfavorable geological conditions as well as large changes in lithology. According to the actual engineering conditions, the SRC of the selected verification area was III, and the lithology was sandstone. As shown in Fig. 9, the mileage ranges of the tunneling test section were 0+598 to 0+561. The range of the experimental contrast section was 0+790 to 0+708. The verification process was as follows: first, the main control parameter value under the TBM operators' experience was used to operate in verification section I, as shown in the red area in Fig. 9. Then, the optimal decision method was used to make the decision of the main control parameters. The thrust and torque values obtained from the optimized decision were used for tunneling in the engineering verification section II, which is represented by the green area in Fig. 9. The effectiveness of the method was verified by comparing the control parameters selected by the operator and the control parameters obtained using the optimal decision as well as the PR, $E_f$, and cost of the two sections.

**Table 10**
Geological condition and average values of main rock mass parameters.

| UCS | SRC | CI | M (mm) | $M_{gt}$ | RQD | CAI | q |
|---|---|---|---|---|---|---|---|

| (MPa) | | | | | (%) | | (%) |
|---|---|---|---|---|---|---|---|
| 78.43 | III | 432.92 | 12.69 | 2/3/4 | 35.17 | 3.28 | 75.14 |

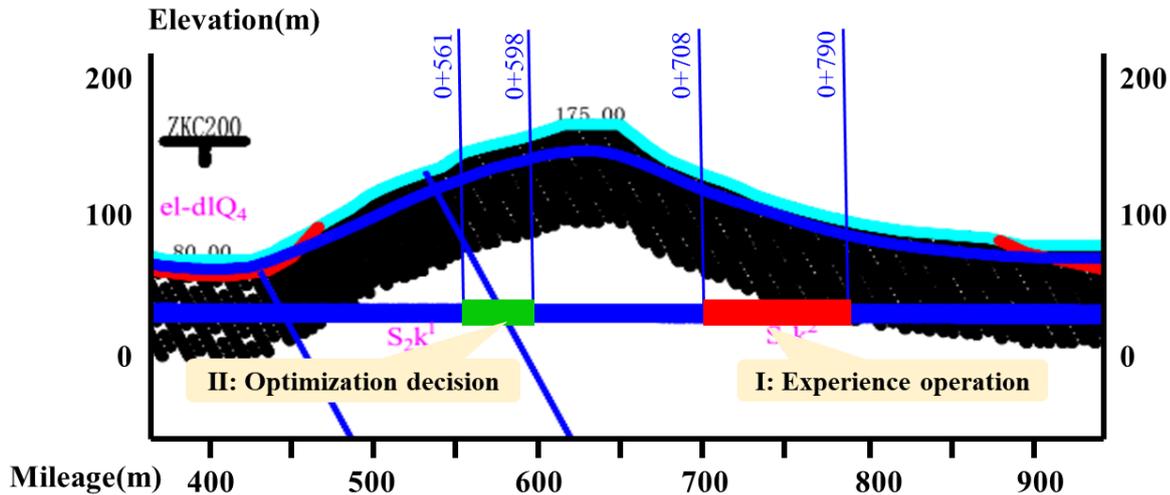

**Fig. 9** Engineering geological profile.

Through the previous geological exploration and analysis of relevant data, the geological rock mass conditions in Section II were relatively stable. We obtained the rock mass mechanical parameters of three locations and selected the average value as the input rock mass parameters of this section, as shown in Table 10. There were three types of muck geometry involved; thus, the main control parameters of the engineering verification section were optimized according to the different muck geometry types. According to the statistics on the construction site, the average cost of each cutter was 30000 RMB. Hence, $c_1$ was 30000. The total cost of the mechanical equipment and labor of the project was approximately 350000 RMB per day. Therefore, $c_2$ was 350000. In the actual operation of the TBM, the range of thrust and torque selected in this study depended on the historical data in normal tunneling of the excavated section of this project. The thrust ranged from 2105.16–9127.08 KN. The torque ranged from 222.49–1327.25 KN·m. Therefore, in the TBM main control parameter optimization decision model, the thrust interval was set as 2000–10000 KN, torque range was set as 200–1500 KN·m. When setting the step size, if the value is too large, the result was not representative, and the best point cannot be found. If the step size is set too small, it is difficult to accurately control in practice, and the amount of calculation increases sharply; thus, the real-time application is limited. As a result, the thrust step was set to 100 KN and the torque step was 50 KN·m. More than 2000 sets of thrust and torque parameters and corresponding cost data were established in the preset interval with a specified step length.

In Section II, the control parameters obtained from the optimized decision were used for tunneling. Except for the different types of muck geometry in the input parameters, the rest of the rock mass parameters were the same. Therefore, three types of thrust and torque values were obtained by optimization. At the same time, the average value of thrust and torque corresponding to each muck geometry type during section I of tunneling by the TBM operator were obtained. The comparison results of the control parameters are shown in Table 11. It can be seen from the table that after the optimization decision, the corresponding control parameters when the cost was optimal were higher than those used by TBM operators in their experience. The relationship model between the cost and the control parameters is shown in Fig. 10; the position of the optimal cost being marked with a red circle in the figure.

Meanwhile, in engineering sections I and II, PR, $E_f$, and cost corresponding to each muck geometry type were also determined, as shown in Table 12. It can be seen from the table that the PR and $E_f$ had increased, and the cost had decreased. In the condition where the muck geometry type was rock debris and rock slices, the PR increased to 68.04 mm/min and $E_f$ increased to 45.21 m$^3$/mm. Compared with the PR and $E_f$ relying on the experience of TBM operators, they increased by 12.61% and 17.03%, respectively. The cost decreased from 10532.50 RMB to 9323.98 RMB, a decrease of 11.47%. In the condition where the muck geometry type was rock debris and rock block, the PR and $E_f$ increased by 14.48% and 15.54%, respectively, reaching 70.74 mm/min and 40.22 m$^3$/mm. Under the condition of $M_{gt}$ 4 (rock debris, rock slices and rock block), the increase in the PR was small, at 6.5%, an increase to 68.98 mm/min. The $E_f$ increased by 32.03 m$^3$/mm, and the decrease in cost was also smaller, dropping by 6.84% to 9515.31 RMB, reducing the cost by 12.73%. In the entire tunneling test section, the average increases in the PR and $E_f$ were 11.10% and 15.62%, respectively, and the average cost decreased by 10.37%. The comparison results are shown in Table 13. The results show that the main control parameter optimization decision method could play an effective role in TBM tunneling in this test section. From the perspective of change trends, it shows that the use of the optimization decision main control parameters in engineering applications could improve the PR and reduce the wear degree of the disc cutter. Moreover, it could improve the overall construction efficiency and reduce costs.

However, the optimization and decision method of the TBM control parameter proposed in this paper also exhibited a few limitations. To enable the method to be more accurate and practical, we will need to continue to explore the following aspects in future research. First, the strength, abrasiveness, quartz content, and other parameters of the rock mass in this study were obtained by mechanical experiments, which lack real-time data. At present, the processing of muck information is mainly based on qualitative analysis, which lacks an accurate and effective quantitative analysis method. In this study, a simple classification form was used for the muck geometry types. In addition, because of the need for conducting a sieving experiment, the CI and M of the muck in tunneling test were set according to the average value. This also has limitations for muck that does not have a fixed shape. Consequently, an in-situ testing robot will be developed to extract effective rock mass parameters dynamically in the future to ensure the real time and effectiveness of parameter acquisition in the multi-objective dynamic optimization method. Further, it will carry out relevant research on real-time acquisition of rock muck information. Second, in the establishment of the rock–machine interaction relationship model, a machine learning algorithm was used to mine rock mass and machine parameters in this study. Without considering the mechanical process and mechanism of the TBM rock breaking and tunneling, its universality is limited. Therefore, in future studies, the rock–machine relationship model driven by physical rules and data statistics will be formed by combining linear cutting test. Finally, with respect to optimization target selection, the optimization objective of this study is the optimal cost, without considering the energy consumption in TBM tunneling. In later research, energy consumption can be included in the scope of cost to optimize the control parameters. There is still more work to be done in the future to explore a more scientific and reasonable intelligent decision method for TBM tunneling.

**Table 11**

Empirical and optimization decision values of control parameters in section I and II.

| Muck geometry types | Control parameters | Operator's experience values | Optimization decision values |
|---|---|---|---|
| Rock debris and rock slices | Th (KN) | 6183.67 | 8600 |
| | Tor (KN·m) | 749.67 | 1350 |
| Rock debris and rock block | Th (KN) | 5068.45 | 5900 |
| | Tor (KN·m) | 780.72 | 950 |
| Rock debris, rock slices and rock block | Th (KN) | 6201.41 | 7500 |
| | Tor (KN·m) | 861.32 | 1350 |

**Table 12**

PR, $E_f$, and cost in section I and II before and after optimization.

| Muck geometry types | | Before optimization | After optimization | Rate of change (%) |
|---|---|---|---|---|
| Rock debris and rock slices | PR (mm/min) | 60.42 | 68.04 | 12.61 |
| | $E_f$ (m$^3$/mm) | 38.63 | 45.21 | 17.03 |
| | Cost (RMB) | 10532.50 | 9323.98 | 11.47 |
| Rock debris and rock block | PR (mm/min) | 61.79 | 70.74 | 14.48 |
| | $E_f$ (m$^3$/mm) | 34.81 | 40.22 | 15.54 |
| | Cost (RMB) | 10414.78 | 9089.32 | 12.73 |
| Rock debris, rock slices and rock block | PR (mm/min) | 64.75 | 68.98 | 6.5 |
| | $E_f$ (m$^3$/mm) | 28.14 | 32.03 | 13.82 |
| | Cost (RMB) | 10214.12 | 9515.31 | 6.84 |

**Table 13**
Comparison of average values before and after optimization decision.

|  | Before decision | After decision | Rate of change |
|---|---|---|---|
| PR (mm/min) | 62.32 | 69.24 | 11.10% |
| $E_f$ (m$^3$/mm) | 33.86 | 39.15 | 15.62% |
| Cost (RMB) | 10387.13 | 9309.54 | 10.37% |

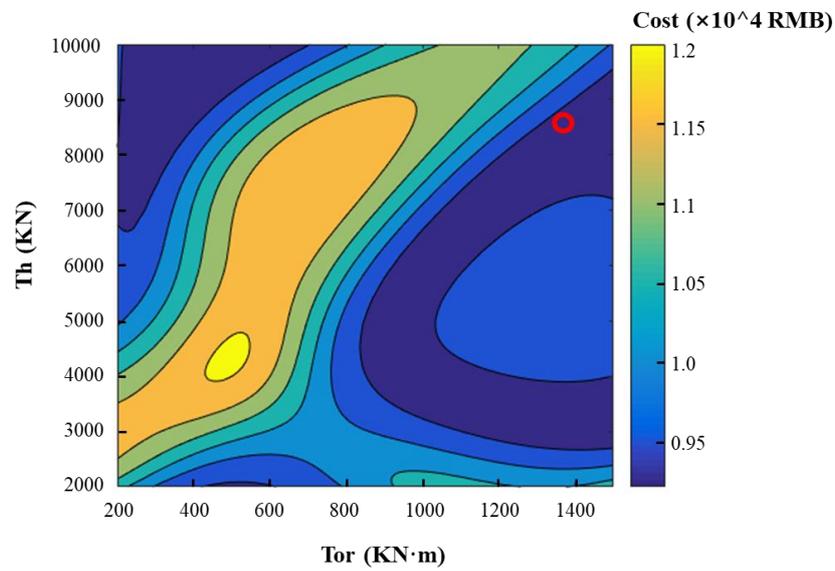

(a)

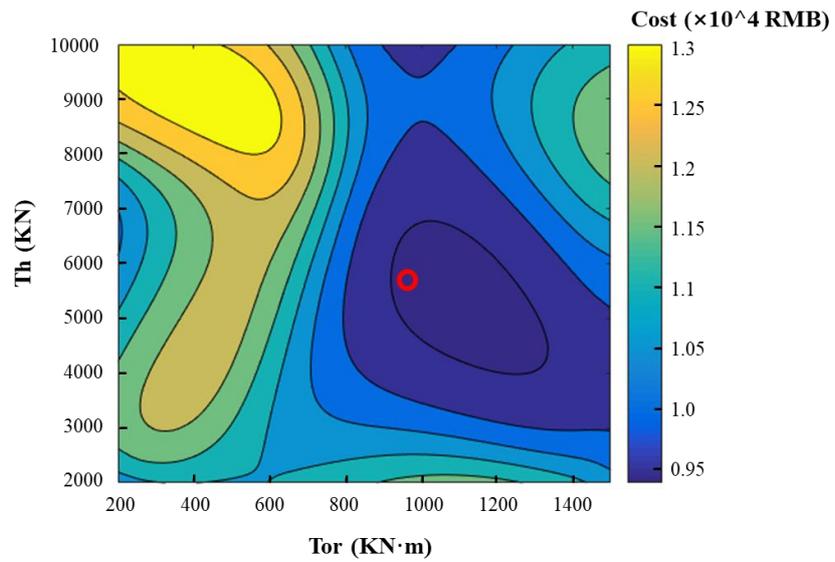

(b)

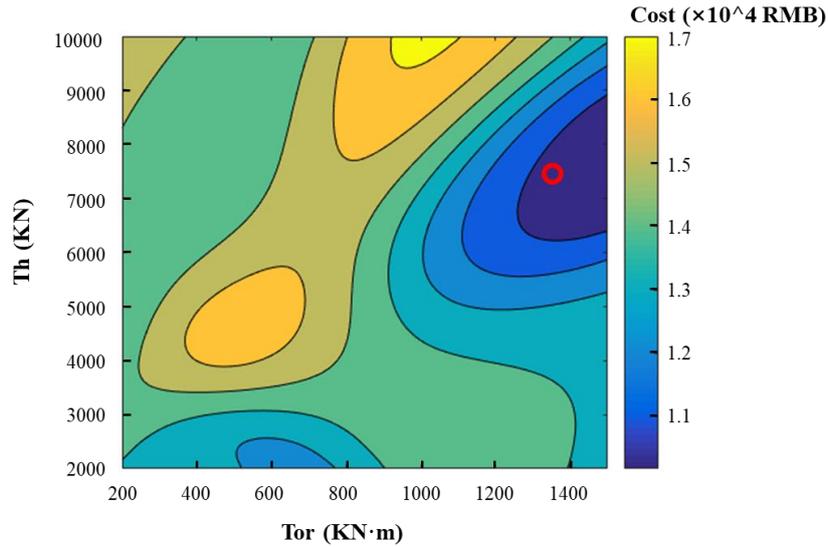

(c)

**Fig. 10** The relationship model between the cost and the control parameters. The muck geometry type represented by (a) is: rock debris and rock slices; (b) is rock debris and rock block; (c) is rock debris, rock slices and rock block. The red circle is the optimal cost.

## 7. Conclusions

This study proposes an intelligent decision method for the TBM main control parameters based on multi-objective optimization of excavation efficiency and cost. It provides an evaluation standard for the selection of TBM control parameters. The main conclusions of this research can be summarized as follows:

1) The muck and cutter wear information were used as important parameters for establishing the rock–machine interaction model as they provide more abundant data support for establishing more accurate rock–machine mapping. The accuracy of the model calculation was improved.

2) The PRCR and CCR models were established using the SA-BPNN algorithm. The MAE and MAPE obtained from the test set of PRCR model were 8.89 mm/min and 13.14%, respectively. The MAE and MAPE obtained from the test set of CCR model were 4.74 m$^3$/mm and 14.06%, respectively. Compared with the traditional BPNN algorithm, the accuracy of the SA-BPNN algorithm is improved. These two models were the basis for the decision method of the main control parameters.

3) The cost calculation model was established using the variables of PR and $E_f$. The decision method was proposed with the goal of optimizing the cost. In the field tunneling test, compared with the section where TBM operators rely solely on experience, the average PR and $E_f$ of the optimized decision tunneling section were significantly improved. The results show that the TBM control parameter optimization decision method proposed in this study is feasible and effective for TBM intelligent tunneling in the test section. However, this method has a few limitations as well. We will continue to further explore these aspects in our future studies.

## Acknowledgements

The authors wish to thank Zhejiang Tunnel Engineering Group Co., Ltd., China Railway Engineering Equipment Group Co., Ltd., and their project designers and constructors for sharing their experiences on data gathering in the field and strong support in the tunneling test. This research was supported by the National Natural Science Foundation of China (NSFC) (No. 51739007), National Science Fund for Excellent Young Scientists Fund (No. 51922067), Joint Program of the National Natural Science Foundation of China (No. U1806226), National Natural Science Foundation of China (NSFC) (No. 51991391), and Taishan Scholars Program of Shandong Province of China (tsqn201909003).